\documentclass[aps,prb,twocolumn,superscriptaddress,floatfix,showpacs]{revtex4}
\usepackage{graphicx}
\usepackage{amssymb}
\usepackage{amsmath}
\usepackage{verbatim}

\begin{document}

\title{Sequential quantum-enhanced measurement with an atomic ensemble}

\author{A.V.\ Lebedev}
\affiliation{Theoretische Physik, Wolfgang-Pauli-Strasse 27, ETH Zurich,
CH-8093 Z\"urich, Switzerland}

\author{P.\ Treutlein}
\affiliation{Department of Physics, University of Basel, Klingelbergstrasse
82, CH-4056 Basel, Switzerland}

\author{G.\ Blatter}
\affiliation{Theoretische Physik, Wolfgang-Pauli-Strasse 27, ETH Zurich,
CH-8093 Z\"urich, Switzerland}

\date{\today}

\begin{abstract}

We propose a quantum-enhanced iterative (with $K$ steps) measurement scheme
based on an ensemble of $N$ two-level probes which asymptotically approaches
the Heisenberg limit $\delta_K \propto R^{-K/(K+1)}$, $R$ the number of
quantum resources.  The protocol is inspired by Kitaev's phase estimation
algorithm and involves only collective manipulation and measurement of the
ensemble. The iterative procedure takes the shot-noise limited primary
measurement with precision $\delta_1\propto N^{-1/2}$ to increasingly precise
results $\delta_K\propto N^{-K/2}$.  A straightforward implementation of the
algorithm makes use of a two-component atomic cloud of Bosons in the precision
measurement of a magnetic field.

\end{abstract}


\maketitle

Estimating an unknown parameter $\phi$ of a quantum system usually involves a
quantum probe prepared in a known state $\hat\rho_0$ which, when brought into
interaction with the system, evolves to a new state $\hat\rho_\phi$ under the
action of a (so-called) quantum channel\cite{giovannetti11} ${\cal Q}_\phi$,
$\hat\rho_0 \rightarrow {\cal Q}_\phi(\hat\rho_0) = \hat\rho_\phi$; by
measuring a suitable observable of the probe in the state $\hat\rho_\phi$, one
can infer the value of $\phi$.  Due to the fundamental uncertainty of the
quantum measurement, this information has a statistical character. In order to
improve the know\-ledge on $\phi$, one needs to repeat the measurement:
measuring $R \gg 1$ independent quantum probes $\hat\rho_0^{\otimes R}
\rightarrow \bigl[{\cal Q}_\phi(\hat\rho_0)\bigr]^{\otimes R} =
\hat\rho_\phi^{\otimes R}$ leads to a $\sqrt{R}$ increase in the measurement
precision $\delta_0 \rightarrow \delta_0/ \sqrt{R}$---this is the standard
quantum limit or shot-noise limit of measurement.  Incidentally, nature
provides us with a better, albeit ultimate\cite{berry12}, Heisenberg limit
$\delta_0 \rightarrow \delta_0/R$, if one exploits some of the quantum
resources of the system under investigation.  Exploiting such quantum effects
enhancing the measurement precision is the subject of quantum metrology
\cite{helstrom76,giovannetti11}.

The quantum enhancement in the measurement precision can be approached by
using either {\it parallel} or {\it sequential} strategies\cite{giovanetti06}.
In a parallel strategy, the original ensemble is divided into $m$
sub-ensembles with $n$ probes, $R = nm$, with each sub-ensemble prepared in a
 (maximally) entangled state
$\bigl[\hat\rho_0^{(n)}\bigr]^{\otimes m} \rightarrow \bigl[{\cal
Q}_\phi(\hat\rho_0^{(n)})\bigr]^{\otimes m}$; this results (ideally) in a
$\sqrt{n}$ enhancement of the precision compared to the standard quantum limit, $\delta_0 \rightarrow \delta_0/
\sqrt{m}\, n = \delta_0\sqrt{m}/R$, see Refs.\ \onlinecite{caves81} for the
case of quantum interferometry and Refs.\ \onlinecite{fujiwara01} for a
general quantum channel.  Alternatively, in a sequential strategy, instead of
preparing entangled states, each of $m$ separate probes is passed $n$ times
through the same quantum channel, $\hat\rho_0^{\otimes m} \rightarrow
\bigl[{\cal Q}_\phi^n(\hat\rho_0) \bigr]^{\otimes m}$, resulting in the same
enhancement in precision, see Refs.\ \onlinecite{luis02}. Roughly speaking,
while parallel strategies make use of  entanglement, sequential
strategies exploit the coherent quantum dynamics as a
resource in order to enhance  measurement precision. While the sequential
strategies do not require creation of rather fragile entangled states, they do
demand longer (coherence) times to allow for the completion of the
measurement.

Recently, quantum measurement protocols were implemented using
two-component atomic ensembles with $N$ particles
\cite{treutlein_10,Ockeloen2013,experiments}. In Ref.\ \onlinecite{treutlein_10},
interatomic interactions were used to create entangled, spin-squeezed
states~\cite{ueda93} of Bose-Einstein condensates (BECs) that allow one to go beyond the standard
quantum limit via a parallel strategy. Here, we suggest a sequential strategy
without entanglement allowing to reach a given precision $\delta$ within
$K\sim \ln(\delta_1/\delta)/\ln(\sqrt{N})$ steps, where $\delta_1=\delta_0/\sqrt{N}$ is the uncertainty of the primary measurement and $N\gg 1$ is the BEC atom number. 
This is achieved by subjecting the BEC on each subsequent measurement step to a $\sim\sqrt{N}$-fold longer evolution under the unknown external field. Our approach resembles Kitaev's phase estimation
algorithm\cite{kitaev96}, however, it requires far less steps to complete the
measurement due to our exploiting the large ensemble $N$ of individual probes
in the atom cloud.  Our approach does not require any separate access to a
particular atom in the BEC and relies only on collective manipulations and
measurements.

{\it Primary measurement.} We wish to estimate the real angle (or phase
\cite{prior_know}) $\phi \in [-\pi,\pi]$ in an unknown unitary rotation
$\hat{U}_z[\phi] = \exp[-i\hat\sigma_z\phi/2]$ given an ensemble (with $N \gg
1$ probes) of spin-$1/2$ systems (qubits) and using only {\it collective}
unitary operations over the ensemble.  We use Ramsey interferometry as the
primary measurement and prepare all qubits in the $\hat\sigma_x = 1$ polarized
state $(|\!\!\uparrow\rangle + |\!\!\downarrow\rangle)/ \sqrt{2}$. Applying
 $\hat{U}_z[\phi]$ to the ensemble results in
the state $[{(e^{-i\phi/2}|\!\uparrow\rangle + e^{+i\phi/2} |\!\downarrow
\rangle)} /\sqrt{2}]^{\otimes N}$.  A $\hat{U}_y[-\pi/2]$ pulse rotates the
ensemble into the read-out state $[\cos(\phi/2)\,{|\!\uparrow\rangle} + i
\sin(\phi/2)\,|\!\downarrow\rangle]^{\otimes N}$ and measuring the total
polarization $\hat{S}_z = (1/N)$ $\sum_{i=1}^N \hat\sigma_z^{(i)}$, one
arrives at one of the possible outcomes $S_z = [N_+ - N_-]/N$, with
$N_+\in\{0,\dots,N\}$ and $N_- = N - N_+$ the number of qubits observed in the
$\sigma_z = \pm 1$ states.  The probability to observe a particular value
$\tilde{S}_z$ (we denote by $\tilde{X}$ a realized value of the random
variable $X$) is given by the Bernoulli distribution
\begin{equation}
   {\cal P}(S_z=\tilde{S}_z|\phi) =  \frac{N!}{\tilde{N}_+!\tilde{N}_-!}
   \Bigl( \cos^2\frac\phi2 \Bigr)^{\tilde{N}_+}
   \Bigl( \sin^2 \frac\phi2\Bigr)^{\tilde{N}_-},
\end{equation}
conditioned on the unknown value of $\phi$. Starting with an unbiased,
homogeneous distribution $P(\phi)$ for the parameter $\phi$ (no {\it a
priori} knowledge on $\phi$), the measurement of a particular value of
$\tilde{S}_z$ allows us to improve our statistical information on $\phi$.
Changing variables $\phi \to p=\cos\phi$ and making use of Bayes' theorem, the
{\it a posteriori} probability distribution $P(p|\tilde{S}_z)$ of $p
\in [-1,1]$ knowing the measured result $\tilde{S}_z$ is enhanced by the
factor ${\cal P}(S_z=\tilde{S}_z|\phi)/{\cal P}(S_z)$, hence, after proper
normalization,
\begin{equation}
   P(p|\tilde{S}_z) = \frac{(N+1)!}{2\tilde{N}_+!\tilde{N}_-!}
   \Bigl( \frac{1+p}2\Bigr)^{\tilde{N}_+} \Bigl(\frac{1-p}2\Bigr)^{\tilde{N}_-}.
   \label{pdist}
\end{equation}
In the following, we consider the limit of large $\tilde{N}_+,~\tilde{N}_- \gg
1$, where the distribution $P(p|\tilde{S}_z)$ has a sharp peak near
$\tilde{S}_z$; expanding the exponent in $\exp[\tilde{N}_+ \ln(1+p) +
\tilde{N}_- (1-p)]$ around the maximum, the distribution $P(p|\tilde{S}_z)$
then can be replaced by the normal distribution
\begin{equation}
   P(p|\tilde{S}_z)\! =\! \frac1{\sqrt{2\pi}\sigma}
     \exp\biggl[ -\frac{(p-\tilde{S}_z)^2}{2\sigma^2} \biggr],
     ~~ \sigma^2 = \frac{1-\tilde{S}_z^2}N,
   \label{cos_dist}
\end{equation}
or in another notation, $p \sim {\cal N}(\tilde{S}_z,\sigma^2)$. The overall
result of the above ensemble measurement is summarized in the following
statistical statement: Given the tolerance level $\beta \ll 1$, the precision
of the estimate $p \approx \tilde{S}_z$ is given by
\begin{equation}\label{eq:tol_beta_p}
   \mbox{Prob}\bigl[|p-\tilde{S}_z| \leq g(\beta) \sigma\bigr] = 1-\beta,
\end{equation}
with $g$ determined by $1-\beta = \mbox{erf}(g/\sqrt{2})$ and $\mbox{erf}(x)
= (1/\sqrt{\pi})\int_{-x}^x dt\, e^{-t^2}$ is the standard error function.
Finally, the distribution (\ref{cos_dist}) for $p=\cos\phi$ provides us with
the distribution function for the angle $\phi$,
\begin{equation}
   P(\phi|\tilde{S}_z) = \frac{|\sin\phi|}{2\sqrt{2\pi}\,\sigma}\,
   \exp\biggl[ -\frac{(\cos\phi-\tilde{S}_z)^2}{2\sigma^2}\biggr].
   \label{phi_dis1}
\end{equation}

{\it Complementary measurement}. Since $\cos\phi$ is even in $\phi$, the
estimate for $\phi$ is distributed among two symmetric intervals $|\phi \pm
\tilde\phi| \leq g/\sqrt{N}$ with $\tilde\phi = |\arccos(\tilde{S}_z)|$.
Expanding (\ref{phi_dis1}) near $\phi \approx \pm\tilde{\phi}$, the
distribution of $\phi$ is given by the sum of two normal distributions,
\begin{equation}
   \phi\sim \frac12\sum_{\alpha = \pm 1} {\cal N}(\alpha\tilde\phi,\sigma_1^2),
   \qquad \sigma_1 = \frac{1}{\sqrt{N}},
   \label{phi1}
\end{equation}
describing two equiprobable alternatives $\alpha=\pm 1$ for the angle $\phi$
to be located near $\alpha\tilde\phi$.  In order to distinguish between the
two alternatives of the primary measurement, we perform a second test by
preparing the ensemble in the $\sigma_y = +1$ polarized state, $\bigl([
|\!\uparrow \rangle + i |\!\downarrow \rangle]/\sqrt{2}\bigr)^{\otimes
N^\prime}$. Repeating the Ramsey measurement, the second estimate for the
parameter $p^\prime = \sin\phi$ should result in either $+\sin\tilde\phi$ or
$-\sin\tilde\phi$, thus distinguishing the alternatives $+\tilde\phi$ and
$-\tilde\phi$ provided by the first measurement. Specifically, given the
probability to observe a qubit in the $\sigma_z = \pm 1$ state conditioned on
the result $\alpha \tilde \phi$ of the primary measurement
\begin{eqnarray} \nonumber
  {\cal P}(\hat{\sigma}_z=\pm1| \alpha\tilde\phi)
  &=& \frac{1}{2}\int d\phi\, \frac{1\pm\sin\phi}{2}
   \sum_{\alpha = \pm 1} {\cal N}(\alpha\tilde\phi,\sigma_1^2)
   \\ \label{eq:P_pm_alpha}
   &\approx& \bigl[ 1 \pm \bigl( \alpha\sin\tilde\phi
   - {\sin^2\tilde\phi}\>/{2N}\bigr) \bigr]/2,
\end{eqnarray}
we find that the total polarization of the complementary ensemble is given by
the sum of two normal distributions, $S_z' \sim \sum_\alpha {\cal N}
(S'_{z\alpha},\sigma^{\prime2})/2$, with mean and variance
\begin{equation}
   S'_{z\alpha} \approx \alpha\sin\tilde\phi - \frac{\sin^2\tilde\phi}{2N},
   \quad
  \sigma^{\prime 2} \approx \frac{\cos^2\tilde\phi}{N^\prime}.
\end{equation}
To construct an unbiased classification rule we define the regions
$E_- =\{S_z'|S_z'< \bar{S}'_z\}$ and $E_+ = \{S_z'|S_z'> \bar{S}'_z\}$ with
the boundary $\bar{S}'_z$ set by the condition
\begin{equation} \label{eq:bound}
   {\cal P}(E_-) \equiv \int_{-\infty}^{\bar{S}'_z} dS_z' \, P(S_z')
                 = {\cal P}(E_+) = 1/2.
\end{equation}
In our symmetric situation, $\bar{S}'_z =(S'_{z-}+S'_{z+})/2$. Given a
measured $\tilde{S}_z'$ we then assign the value $\alpha = 1$ ($\alpha = -1$)
whenever the event has been realized in $E_+$ ($E_-$).  This assignment is
prone to a {\it misclassification} error $\beta' = {\cal P}(+|E_-){\cal
P}(E_-)+{\cal P}(-|E_+){\cal P}(E_+)$, where the conditional probability
${\cal P}(+|E_-) = {\cal P}(E_-|+) {\cal P}(+)/{\cal P}(E_-)$ (and similar for
${\cal P}(-|E_+)$) follows from Bayes' theorem; for our unbiased
classification rule, this reduces to ${\cal P}(\alpha'|E_{\alpha}) = {\cal P}
(E_{\alpha} |\alpha')$. The conditional probabilities ${\cal
P}(E_\alpha|\alpha')$ are easily obtained from the distributions ${\cal
P}(S_z'|\alpha) = {\cal N} (S'_{z\alpha}, \sigma^{\prime 2})$ and we find that
\begin{eqnarray}
   \beta^\prime &=&
   [1-\mbox{erf}(|\sin\tilde\phi|/\sqrt{2}\sigma')]/2.
   \label{eq:betap}
\end{eqnarray}
Away from the immediate vicinity of $\tilde\phi \approx 0$ (e.g., $\tilde{N}_+
> 5$) and a typical number  of probes $N \sim 10^3$, choosing $N' \sim N$ results
in a negligible probability $\beta'$ of misclassification.

{\it n-fold rotation.} Next, we analyze the sequential ($n > 1$-fold)
application of the rotation $\hat{U}_z[\phi]$ in the Ramsey measurement.
Measuring the ensemble polarization $\hat{S}_{z}$ then provides an estimate
for the parameter $\phi_n = n\phi$. Given a measured result $\tilde{S}_{zn}$,
the {\it a posteriori} distribution function for the parameter $\phi_n$ is
given by Eq.\ (\ref{phi_dis1}) with $\tilde{S}_{z}\rightarrow \tilde{S}_{zn}$
and $\phi\rightarrow \phi_n$, providing two Gaussian peaks at $\tilde\phi_n =
\pm |\arccos( \tilde{S}_{zn} )|$ of width $\propto 1/\sqrt{N}$ as a function
of $\phi_n$; the accompanying complementary measurement again selects one
alternative $\alpha=\pm 1$.

A further estimate of the angle $\phi$ is obtained by reading the distribution
function Eq.\ (\ref{phi_dis1}) with $\cos\phi \rightarrow \cos( n\phi)$ as a
function of $\phi$. The periodicity of $\cos(n\phi)$ provides $n$ different
values of $\phi$, all corresponding to the same value of $\cos(n\phi)$. The
distribution function $P_n(\phi)$ for the angle $\phi$ then has $n$
peaks centered at $\tilde\phi_{nk}$,
\begin{eqnarray} \label{nfolddist}
   \phi &\sim& \frac1n\sum_{k=0}^{n-1} {\cal N}
   \bigl(\tilde\phi_{nk},\sigma_n^2\bigr), \quad \sigma_n = \frac{\sigma_1}{n},\\
   \tilde\phi_{nk} &=& \frac{\tilde\phi_n}{n} + 2\pi \frac{k}{n},~~~k = 0,\dots,n-1,
\end{eqnarray}
each peak $n$ times narrower than in the previous $n=1$ case.  As a result,
the $n$-fold measurement redistributes the original uncertainty $\delta_1 \sim
\sigma_1$ among $n$ different equiprobable positions which we call the
alternatives $A_k$. While this result does not give us more {\it a posteriori}
information on the position of $\phi$ than the 1-fold measurement (as
confirmed by the Shannon entropy $H_n = - \int d\phi\, P_n(\phi)\ln P_n(\phi)$
coinciding for all $n$), the different distribution of the probability allows
us to gain in precision when combining the two measurements.

\begin{figure}
\includegraphics[width=7.0cm]{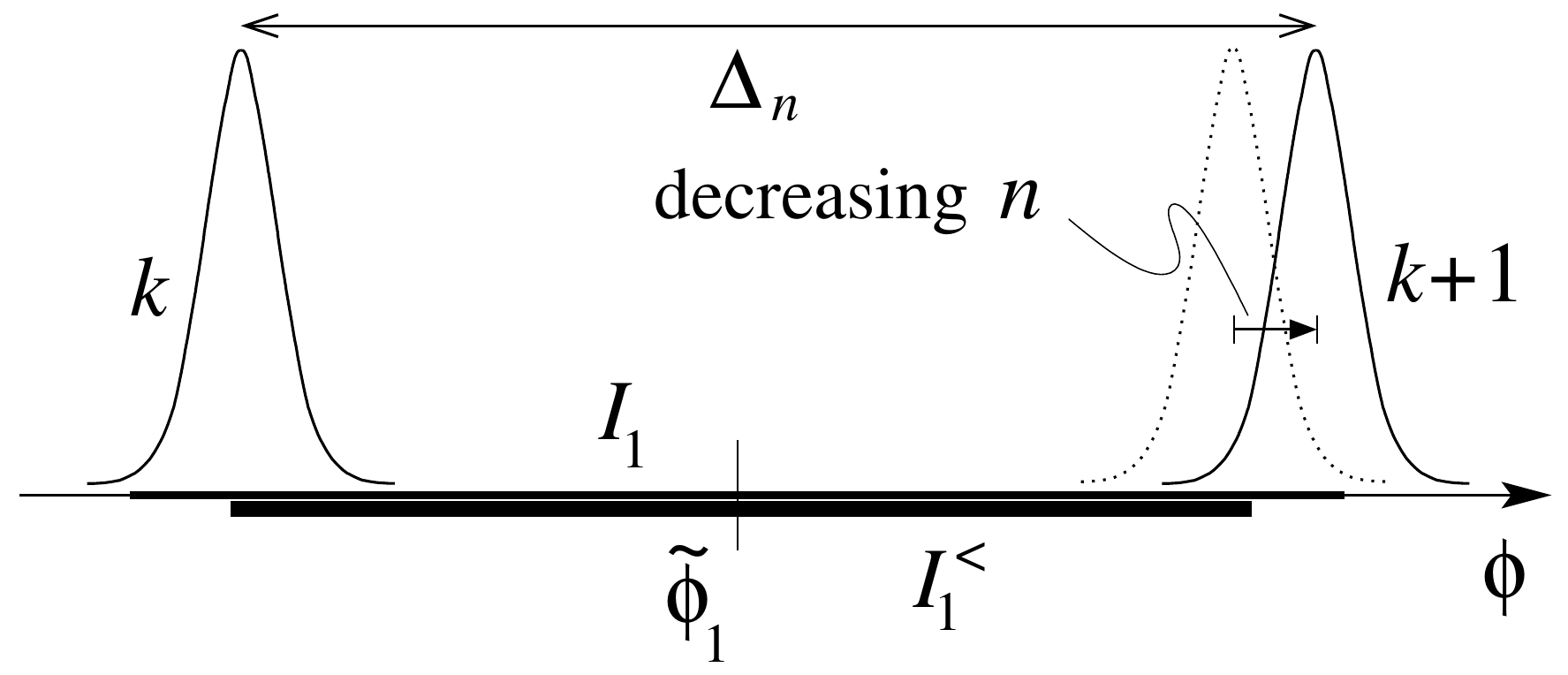}
\caption[]
{\label{fig:int} Intervals $I_1$ and $I_1^{\scriptscriptstyle <}$ centered
around $\tilde\phi_1$. The peaks belonging to the alternatives $A_k$ and
$A_{k+1}$ are shown for the extreme case where $\tilde\phi_{nk}$ coincides
with the left boundary and $\tilde\phi_{nk}$ is on the verge of leaving
$I_1^{\scriptscriptstyle <}$ as $n$ decreases, $\Delta_n = 2\pi/n$.}
\end{figure}

{\it Beyond the shot noise limit.} In order to take advantage of the $n$-fold
measurement one has to identify the correct alternative among the $n$
equiprobable distributions, see  Eq.\ (\ref{nfolddist}). This is done by
combining the results of the 1- and $n$-fold measurements. We define the
interval $I_1 = \{\phi | \> |\phi-\tilde\phi_1|\leq g(\beta)\sigma_1\}$
centered around the result $\tilde\phi_1$ of the first measurement, see Fig.\
\ref{fig:int}, with $\beta$ the tolerance level of the first measurement,
hence $\mbox{Prob}[|\phi-\tilde\phi_1| \leq g(\beta)\sigma_1] = 1-\beta$, cf.\
Eq.\ (\ref{eq:tol_beta_p}). We call the alternative $A_k$ compatible with the
first measurement if the condition
\begin{equation}
   \mbox{Prob}\bigl[ \phi\in I_1 |
   \phi \sim {\cal N}\bigl(\tilde\phi_{nk},\sigma^2_n\bigr) \bigr]
   \geq 1-\beta \label{eq:alt}
\end{equation}
is satisfied. In order to satisfy the original confidence level in the second
measurement, the maximum $\tilde\phi_{nk}$ belonging to $A_k$ must be located
within the reduced interval $I_1^{\scriptscriptstyle <} = \{\phi | \>
|\phi-\tilde\phi_1|\leq g(\beta)(\sigma_1-\sigma_n)\}$, see Fig.\
\ref{fig:int}.

By construction, the condition (\ref{eq:alt}) is satisfied for at least one
$k$, irrespective of the value of $n$. Choosing a small $n$, the gain in
precision is small, hence we are interested in maximizing the value of $n$.  On
the other hand, for large $n$, the number of peaks compatible with Eq.\
(\ref{eq:alt}) is larger than one and we cannot select the proper alternative
$A_k$. The optimal number $n_\mathrm{opt}$ can be determined by considering
the situation where the $k$-th peak is located at the left boundary of
$I_1^{\scriptscriptstyle <}$, $\tilde\phi_{nk} = \tilde\phi_1 -
g(\beta)(\sigma_1-\sigma_n)$, while the next peak $\phi_{nk+1}$ is being
pushed out from $I_1^{\scriptscriptstyle <}$ across the right boundary of
$I_1^{\scriptscriptstyle <}$ with decreasing $n$, see Fig.\ \ref{fig:int}.
Obviously, when $\tilde\phi_{nk+1} = \tilde\phi_1 + g(\beta)
(\sigma_1-\sigma_n)$ we still have two equally probable alternatives
generating a large misclassification error $\tilde\beta = 1/2$.  The task then
is, to find the largest possible $n$ compatible with a prescribed error
$\tilde\beta \ll 1$.

After two measurements, a 1- and a $n$-fold, the {\it a posteriori}
distribution function for the angle $\phi$ is given by,
\begin{equation}
   P(\phi|\tilde\phi_1,\tilde\phi_n) \propto
   \sum_{k=0}^{n-1} \frac{w_k}{\sqrt{2\pi}\>\sigma_{1,n}}
   \exp\biggl[-\frac{(\phi-\tilde\phi_{1,nk})^2}{2\sigma_{1,n}^2}\biggr],
\end{equation}
with $\tilde\phi_{1,nk} = (\tilde\phi_1 \sigma_n^2 +\tilde \phi_{nk}
\sigma_1^2)/(\sigma_1^2+\sigma_n^2) \approx \tilde\phi_{nk}$, $\sigma_{1,n} =
\sigma_1\sigma_n/\sqrt{\sigma_1^2+\sigma_n^2} \approx \sigma_n$, and $w_k$ is
the {\it a posteriori} probability that the $k$-th alternative has been
realized,
\begin{equation}
   w_k = \exp\biggl[-\frac{(\tilde\phi_1-\tilde\phi_{nk})^2}
      {2(\sigma_1^2+\sigma_n^2)} \biggr].
\end{equation}
For our arrangement $\tilde\phi_{nk} = \tilde\phi_1 - g(\beta) (\sigma_1 -
\sigma_n)$ and $\tilde\phi_{nk+1} = \tilde\phi_1- g(\beta) (\sigma_1-\sigma_n)
+\Delta_n$. Considering only these two peaks, the misclassification error of
the $k$-th alternative is given by $\tilde\beta = w_{k+1}/(w_k + w_{k+1})$.
Solving for $\Delta_n$, we obtain the optimal number of rotations compatible
with $\tilde\beta$, $n \leq n_\mathrm{opt} = \lfloor \nu(\beta,\tilde\beta) /
\sigma_1\rfloor \equiv n_2$ with
\begin{equation}
   \nu = \pi g(\beta)
   \frac{\sqrt{1+2\ln[(1-\tilde\beta)/\tilde\beta]/g^2(\beta)}-1}
   {\ln[(1-\tilde\beta)/\tilde\beta]}.
\end{equation}
In the end, the precision estimate of two measurements with a 1-fold and a
$n_2$-fold rotation is given by
\begin{equation}
  \mbox{Prob}\bigl[ |\phi - \tilde\phi_2| \leq
   g(\beta) \sigma_2 \bigr] = (1\!-\!\beta)(1\!-\!\tilde\beta),
  \label{two_step}
 \end{equation}
where $\tilde\phi_2$ is the measured and selected value of the parameter
$\phi$ on the second step and the error bar is given by $\sigma_2 \equiv
\sigma_{n_2} = \sigma_1/n_2$. Thus the second
measurement improves the precision by the large factor
$\sqrt{N}\nu(\beta,\tilde\beta) \gg 1$, with $\nu(\beta,\tilde\beta)\approx
0.96$ for $\beta=\tilde\beta = 0.01$ (for small $\beta,\tilde\beta \ll 1$,
$\nu \approx (\pi\sqrt{2|\ln\beta|)}/ |\ln\tilde\beta|)
[(1+\ln\tilde\beta/\ln\beta)^{1/2}-1]$.  The $K$-fold iteration of this
procedure will further improve the precision of the measurement.

The above measurement protocol involves two sources of error, the {\it
estimation error} $\beta$ for the angle $\phi$ to lie outside the interval
$|\phi-\tilde\phi_2|\leq g(\beta)\sigma_2$, and the {\it classification error}
$\tilde\beta$ for an incorrect choice of the alternative $A_k$. While an
estimation error can be ruled out on a subsequent step, this is not the case
for a classification error. Indeed, for $K=2$ and assuming that we have
correctly identified the peak $\tilde\phi_{nk} = \tilde\phi_2$, let us suppose
that $|\phi-\tilde\phi_2|>g(\beta)\sigma_2$, hence the true value $\phi$ is
outside the allowed range. Applying $n_3 = \nu/\sigma_2$ rotations in the next
step, none of the peaks $\tilde\phi_{n_3 k}$ will belong to the interval
$\tilde{I}_2$ defined through $|\phi-\tilde\phi_2|\leq g(\beta) (\sigma_2 -
\sigma_3)$ with probability $1-\beta$, signalling the error in one of the
previous steps.  The classification error cannot be caught in subsequent
steps: Consider two alternative angles $\phi$ localized either near $a$) the
$k$-th peak $|\phi -\tilde\phi_{n_2k}|\leq g(\beta)\sigma_2$ or $b$) the next
peak $|\phi - (\tilde\phi_{n_2k}+2\pi/n_2)| \leq g(\beta)\sigma_2$.  Then
applying a $n_3$-fold rotation, the random parameter $\phi_3 = n_3\phi$ will
be localized within $a$) $|\phi_3 - n_3\phi_{n_2k}| \leq g(\beta)\nu$, or $b$)
$|\phi_3 -n_3\phi_{n_2k} - 2\pi\nu\sqrt{N}|\leq g(\beta)\nu$.  Since these
intervals are largely overlapping, the hypotheses $a$) and $b$) cannot be
distinguished.

Iterating the process $K$ times, our measurement protocol satisfies the
confidence criterion
\begin{eqnarray}
   \mbox{Prob}\bigl[ |\phi - \tilde\phi_K|
   \leq g(\beta) \sigma_K \bigr] = (1\!-\!\beta)(1\!-\!\tilde\beta)^{K-1},
   \label{kstep}
\end{eqnarray}
with $\sigma_K = \sigma_1 [{\sigma_1} / {\nu}]^{K-1}\sim N^{-K/2}$.  To reach
a given precision $\delta$ one then needs to perform $K_\delta \sim
1+\ln(\sigma_1 g/\delta)/\ln(\nu\sqrt{N})$ steps; at the same time, the
overall confidence level decreases exponentially $\propto \exp(-\tilde\beta
K_\delta)$ with the number $K_\delta$ of steps.  Moreover, an estimation
error, i.e., none of the peaks belongs to the prescribed estimation interval,
forces one to repeat the entire procedure again. However, in practice
$\sqrt{N}\sim 30 - 40$ and thus hardly more than a few steps are required to
reach a good precision $\delta$.

In counting the number of resources needed to reach the estimate
(\ref{kstep}), we choose as our basic unit the operation $\hat{U}_z[\phi]$
applied to a single qubit. The $i$-th step of the above procedure requires
(for $N$ qubits in the ensemble) $R_i = N n_i = N ({\nu}/{\sigma_1})^{i-1} =
\nu^{i-1}\,N^{(i+1)/2}$ elementary operations. For $\nu/\sigma_1 \gg 1$, the
entire $K$-step process then involves $R = \sum_{i=1}^{K} R_i \approx R_{K}$
operations.  Expressing the prescribed precision $\delta_K\> (\,\approx
g\sigma_{K} = g\sigma_1 [{\sigma_1} / {\nu}]^{K-1})$ through the number $R$ of
resources, we arrive at the scaling
\begin{equation}
   \delta_K = \frac{g(\beta)}
                 {\bigl[\nu(\beta,\tilde\beta)\bigr]^{\frac{K-1}{K+1}}}
   \frac{1}{R^{\frac{K}{K+1}}} \sim \frac{\delta_0}{N^{K/2}},
\end{equation}
telling us that the Heisenberg limit is reached asymptotically at large values
of $K$.  In an actual implementation with ultracold atoms, the time needed to
prepare the atomic ensemble is typically a few seconds\cite{treutlein_10}, and
increasing the number of rotations $n$ does not significantly increase the
overall duration of the measurement.  The standard statistical measurement
requires $K_\mathrm{std} \sim (\delta_0/\sqrt{N}\delta)^2$ preparations to
reach a precision $\delta$, which is exponentially larger than the $K_\delta$
preparations required with the present protocol, $K_\mathrm{std} \sim
N^{K_\delta-2}$.

{\it Dephasing.} In addition to the unitary rotation $\hat{U}_z[\phi]$, the
qubits may experience a stochastic field $\varphi(t)$, e.g.\ due to
uncontrolled interactions between qubits generating a different phase shift
$\phi \rightarrow \phi + \int dt\,\varphi(t)$ for each qubit. Averaging the
single-qubit density matrix over $\varphi(t)$, the off diagonal amplitudes
$\epsilon\, \exp(\pm i\phi)$ are reduced by the factor $\epsilon =
\exp(-\Gamma \tau_1 /2)<1$, where we have assumed a Gaussian random field
$\langle \varphi(t) \varphi(t') \rangle = \Gamma \delta(t-t')$ and $\tau_1$ is
the exposure time of the primary measurement; the reduction in these
amplitudes after an $n$-fold rotation is given by $\epsilon^n$. The
measurement of the ensemble polarization involves the parameter $p =
\epsilon^n\cos(n\phi)$ and the width in the {\it a posteriori} distribution
function is $\sigma_n \rightarrow \sigma_1/(n\epsilon^n)$. The smallest
attainable width $\sigma_1 e\, \ln(1/\epsilon)$ is reached after $n_c =
-1/\ln\epsilon = \tau_c/\tau_1$ steps, with $\tau_c=2/\Gamma$ the coherence
time.
%

{\it Application.} We analyze the use of our protocol to measure a constant
magnetic field $B$ with an atomic ensemble, considering a transition with
differential magnetic moment of order $\mu_{\rm \scriptscriptstyle B}$, the
Bohr magneton. Assuming the prior knowledge that $B < B_+$, we choose the
interrogation time of the first Ramsey sequence $\tau_1 \sim 2\pi
\hbar/\mu_{\rm\scriptscriptstyle B} B_+$ such that the accumulated phase $\phi
= \mu_{\rm \scriptscriptstyle B} B \tau_1 /\hbar$ does not exceed $2 \pi$.
This primary measurement results in a phase uncertainty of $[\delta\phi]_1 =
1/\sqrt{N}$, translating to a precision of $[\Delta B]_1 =
\hbar/\mu_{\rm\scriptscriptstyle B} \tau_1 \sqrt{N}$ in the field. In the
following steps, the Ramsey time is increased as described above. The longest
Ramsey sequence of duration $\tau_c = n_c \tau_1$ provides us with a precision
of $[\delta\phi]_\mathrm{min} = 1/n_c\sqrt{N}$ for the phase estimation, and
thus a field precision of $[\Delta B]_\mathrm{min} =\hbar/ \mu_{\rm
\scriptscriptstyle B} \tau_c \sqrt{N}$ using $K \sim 1 + \ln(\tau_c/\tau_1)
/\ln(\sqrt{N})$ steps.

In a realistic situation the above procedure is feasible for small magnetic
fields, since the duration $\tau_1$ of the first Ramsey sequence cannot be
arbitrarily small---a typical $\tau_1 \sim 10^{-6}$ s corresponds to a field
$B_+ \sim 1$ G.  With a typical coherence time $\tau_c \sim 1$ s and $N\sim
1000$ atoms one arrives after $K = 5 $ steps at a precision $\Delta B \sim
3\times 10^{-9}$~G.  In order to measure higher fields one can exploit the
phase periodicity and subtract an offset field. This requires prior knowledge
that the field lies in an interval $[B_-,B_+]$. In this case, given a minimal
time $\tau_1$, we choose some field $B_0 \in[B_-,B_+]$ that satisfies the
matching relation $\mu_{\rm \scriptscriptstyle B} B_0\tau_1/\hbar = 2\pi M$
with $M$ the largest possible integer. The procedure described above is then
used to measure the remaining small field $b = B - B_0$.
%

{\it Conclusion.} It is interesting to compare our ensemble-based algorithm
with Kitaev's original phase estimation algorithm involving individual qubits.
In order to reach a prescribed precision $\delta$, the latter necessitates $K
\sim \ln(1/\delta) / \ln 2$ steps, a factor $(\ln \sqrt{N})/ \ln 2$ larger
than the ensemble-based protocol. For the Kitaev algorithm, the resources
scale as $\delta \sim \ln R / R$ (accounting for the fact that a final error
probability $\beta$ necessitates a smaller value $\beta/K$ for the individual
step \cite{kitaev96}), which is better than our algebraic relation $\delta
\sim R^{-K/(K+1)}$. Still, performing a few iterations on an ensemble of
$N\sim 10^3$ atoms with long coherence time appears as an attractive and
practical alternative to control isolated qubits over many steps.
%

The sequential strategy discussed here is particularly useful in scanning
probe measurements of a spatially varying field\cite{Ockeloen2013}. If the
field distribution is {\it a priori} unknown, the duration of the initial
Ramsey sequence has to be very short at each pixel of the image. Our scheme
allows to optimally adapt the sequence at each pixel to reach fast reduction
in the measurement uncertainty. 
This dramatically reduces the overall time  to record a picture with a given
precision. Finally, our ensemble-based sequential strategy could also be
combined with a parallel strategy using squeezed BECs.

We acknowledge fruitful discussions with Gordey Lesovik and Tilman Esslinger
and financial support from the Swiss National Foundation through the NCCR QSIT.


\begin{thebibliography}{99}

\bibitem{giovannetti11} V.\ Giovannetti, S.\ Lloyd, and L.\ Maccone, Nature
Photonics {\bf 5}, 222 (2011); V.\ Giovannetti, S.\ Lloyd, and L.\ Maccone,
Science {\bf 306}, 1330 (2004).

\bibitem{berry12} D.W.\ Berry, M.J.W.\ Hall, M.\ Zwierz, and H.M.\
Wiseman, Phys. Rev. A {\bf 86}, 053813 (2012); V.\ Giovannetti, S.\ Lloyd, and
L.\ Maccone, Phys.\ Rev.\ Lett.\ {\bf 108}, 260405 (2012).

\bibitem{helstrom76} C.W.\ Helstrom, {\it Quantum Detection and Estimation
Theory} (Academic Press, New York, 1976).

\bibitem{giovanetti06} V.\ Giovannetti, S.\ Lloyd, and L.\ Maccone, Phys.\ Rev.\
Lett.\ {\bf 96}, 010401 (2006).

\bibitem{caves81} C.M.\ Caves, Phys.\ Rev.\ D {\bf 23}, 1693 (1981); D.J.\
Wineland, J.J.\ Bollinger, W.M.\ Itano, and F.L.\ Moore, Phys.\ Rev.\ A
{\bf 46}, R6797 (1992); M.J.\ Holland and K.\ Burnett, Phys.\ Rev.\ Lett.\ {\bf
71}, 1355 (1993).

\bibitem{fujiwara01} A.\ Fujiwara, Phys. Rev. A {\bf 63}, 042304 (2001); D.G.\
Fischer, H.\ Mack, M.A.\ Cirone, and M.\ Freyberger, Phys.\ Rev.\ A {\bf
64}, 022309 (2001); M.\ Sasaki, M.\ Ban, and S.M.\ Barnett, Phys.\ Rev.\ A
{\bf 66}, 022308 (2002).

\bibitem{luis02} A.\ Luis, Phys. Rev. A {\bf 65}, 025802 (2002); T.\ Rudolph
and L.\ Grover, Phys.\ Rev.\ Lett.\ {\bf 91}, 217905 (2003); M.\ de Burgh and
S.D.\ Bartlett, Phys.\ Rev.\ A {\bf 72}, 042301 (2005); B.L.\ Higgins, D.W.\
Berry, S.D.\ Bartlett, H.M.\ Wiseman, and G.J.\ Pryde, Nature {\bf 450},
393 (2007); W.\ van Dam, G.M.\ D'Ariano, A.\ Ekert, C.\ Macchiavello, and
M.\ Mosca, Phys.\ Rev.\ Lett.\ {\bf 98}, 090501 (2007).

\bibitem{huelga97} S.F.\ Huelga, C.\ Macchiavello, T.\ Pellizzari, and A.K.\
Ekert, Phys.\ Rev.\ Lett.\ {\bf 79}, 3865 (1997); J.\ Koodynski and R.\
Demkowicz-Dobrzanski, Phys.\ Rev.\ A {\bf 82}, 053804 (2010).

\bibitem{treutlein_10} C.\ Gross, T.\ Zibold, E.\ Nicklas, J.\ Est\`eve, and
M.K.\ Oberthaler, Nature {\bf 464}, 1165 (2010); 
M.F.\ Riedel, P.\ B\"ohi, Y.\ Li, T.W.\ H\"ansch, A.\ Sinatra, 
and P.\ Treutlein, Nature {\bf 464}, 1170 (2010).

\bibitem{Ockeloen2013} C.F.\ Ockeloen, R.\ Schmied, M.F.\ Riedel, 
and P.\ Treutlein, arXiv:1303.1313 (2013). 

\bibitem{experiments} A.\ Louchet-Chauvet, J.\
Appel, J.J.\ Renema, D.\ Oblak, N.\ Kjaergaard, and E.S.\ Polzik, New J.\
Phys.\ {\bf 12}, 065032 (2010); I.\ Leroux, M.\ Schleier-Smith, and V.\
Vuletic, Phys.\ Rev.\ Lett.\ {\bf 104}, 250801 (2010).

\bibitem{ueda93} M.\ Kitagawa and M.\ Ueda, Phys.\ Rev.\ A {\bf 47}, 5138
(1993).

\bibitem{kitaev96} A.Y.\ Kitaev, Electr.\ Coll.\ Comput.\ Complex.\ {\bf 3},
(1996); Z.\ Ji, G.\ Wang, R.\ Duan, Y.\ Feng, and M.\ Ying, IEEE Trans.\ Inf.\
Theory {\bf 54}, 5172 (2008).

\bibitem{prior_know} Multiple winding of the phase $\phi$ is excluded,
reflecting some prior knowledge on the unknown field; otherwise, our protocol
estimates $\phi$ only modulo $2\pi$.

\end{thebibliography}
\end{document}